\documentclass[reprints,amsmath,amssymb,superscriptaddress]{revtex4}
\usepackage{graphicx}
\begin{document}

\title{Disentangling the electronic and phononic glue in a high-$T_\mathrm{c}$ superconductor} 

\author{S. Dal Conte}
\email{s.f.p.dalconte@tue.nl}
\affiliation{Department of Physics A. Volta,Universit\`a degli Studi di Pavia, Pavia I-27100, Italy}
\affiliation{Present address: Physics Department, Eindhoven University of Technology, the Netherlands}

\author{C. Giannetti}
\email{claudio.giannetti@unicatt.it}
\affiliation{i-Lamp, Interdisciplinary Laboratories for Advanced Materials Physics}
\affiliation{Department of Physics, Universit\`a Cattolica del Sacro Cuore, Brescia I-25121, Italy}

\author{G. Coslovich}
\affiliation{Department of Physics, Universit\`a degli Studi di Trieste, Trieste I-34127, Italy}
\affiliation{Present address: Materials Sciences Division, Lawrence Berkeley National Laboratories, CA, USA}

\author{F. Cilento}
\affiliation{Department of Physics, Universit\`a degli Studi di Trieste, Trieste I-34127, Italy}

\author{D. Bossini}
\affiliation{Department of Physics, Universit\`a Cattolica del Sacro Cuore, Brescia I-25121, Italy}
\affiliation{Present address: IMM, Radboud University Nijmegen, the Netherlands}

\author{T. Abebaw}
\affiliation{Department of Physics, Universit\`a degli Studi di Trieste, Trieste I-34127, Italy}
\affiliation{Present address: Zernike Institute for Advanced Materials, University of Groningen, the Netherlands}

\author{F. Banfi}
\affiliation{i-Lamp, Interdisciplinary Laboratories for Advanced Materials Physics}
\affiliation{Department of Physics, Universit\`a Cattolica del Sacro Cuore, Brescia I-25121, Italy}

\author{G. Ferrini}
\affiliation{i-Lamp, Interdisciplinary Laboratories for Advanced Materials Physics}
\affiliation{Department of Physics, Universit\`a Cattolica del Sacro Cuore, Brescia I-25121, Italy}

\author{H. Eisaki}
\affiliation{Nanoelectronics Research Institute, National Institute of Advanced Industrial Science and Technology, Tsukuba, Japan}

\author{M. Greven}
\affiliation{School of Physics and Astronomy, University of Minnesota, Minneapolis, Minnesota 55455, USA}

\author{A. Damascelli}
\affiliation{Department of Physics {\rm {\&}} Astronomy, University of British Columbia, Vancouver, Canada}
\affiliation{Quantum Matter Institute, University of British Columbia, Vancouver, Canada}

\author{D. van der Marel}
\affiliation{D\'epartement de Physique de la Mati\`ere Condens\'ee, Universit\'e de Gen\`eve, Switzerland}

\author{F. Parmigiani}
\affiliation{Department of Physics, Universit\`a degli Studi di Trieste, Trieste I-34127, Italy}
\affiliation{Sincrotrone Trieste S.C.p.A., Basovizza I-34012, Italy}

\begin{abstract}
\textbf{Unveiling the nature of the bosonic excitations that mediate the formation of Cooper pairs is a key issue for understanding unconventional superconductivity. A fundamental step toward this goal would be to identify the relative weight of the electronic and phononic contributions to the overall frequency ($\Omega$) dependent bosonic function, $\Pi(\Omega)$. We perform optical spectroscopy on Bi$_{2}$Sr$_{2}$Ca$_{0.92}$Y$_{0.08}$Cu$_{2}$O$_{8+\delta}$ crystals with simultaneous time- and frequency-resolution; this technique allows us to disentangle the electronic and phononic contributions by their different temporal evolution. The strength of the interaction ($\lambda$$\sim$1.1) with the electronic excitations and their spectral distribution fully account for the high critical temperature of the superconducting phase transition.} 
\end{abstract}
\maketitle

Lattice vibrations \cite{Kresin2009} and excitations of electronic origin, like spin or electric polarizability fluctuations\cite{Monthoux2007} and loop currents\cite{Varma2006}, are generally considered potential mediators of Cooper-pairing in the copper-oxide high-temperature superconductors (cuprates). The generic interaction of fermionic quasiparticles (QPs) with bosonic excitations is accounted for by the bosonic function $\Pi(\Omega)$ (usually indicated as $\alpha^2F(\Omega)$ for phonons and $I^2 \chi(\Omega)$ for spin fluctuations), a dimensionless function that depends on the density of states of the excitations and the strength of their coupling  to QPs. Because both the energy dispersion and lifetime of QPs are strongly affected by the interactions, signatures of QP-boson coupling have been observed in experiments that probe the electronic properties at equilibrium. The ubiquitous kinks in the QP dispersion at $\sim$70 meV, measured by angle-resolved photoemission spectroscopy (ARPES)\cite{Damascelli2003}, have been interpreted in terms of coupling to either optical Cu-O lattice modes\cite{Lanzara2001,Devereaux2004} or spin excitations\cite{Dahm2009}. Inelastic neutron and X-ray scattering experiments found evidence for both QP-phonon anomalies\cite{Reznik2006} and bosonic excitations attributed to spin fluctuations\cite{Dahm2009,LeTacon2011} and loop currents\cite{Li2010}. Dip features in tunnelling experiments have been used to alternatively support the scenarios of dominant electron-phonon interactions\cite{Lee2006} or antiferromagnetic spin fluctuations\cite{Ahmadi2011}. The frequency-dependent dissipation of the Drude optical conductivity, $\sigma\left(\omega\right)$, measured by equilibrium optical spectroscopies, has been interpreted\cite{Norman2006,Hwang2007,vanheumen2009} as the coupling of electrons to bosonic excitations, in which the separation of the phononic and electronic contributions is impeded by their partial coexistence on the same energy scale ($<$90 meV). 
\\

We disentangle the electronic and phononic contributions to $\Pi(\Omega)$ through a non-equilibrium optical spectroscopy, in which the femtosecond time-resolution is combined with an energy-resolution smaller than 10 meV, over a wide photon energy range (0.5-2 eV).
Our approach is based on the widely-used assumption\cite{Perfetti2007,Carbone2008} that, after the interaction between a superconductor and a short laser pulse (1.55 eV photon energy), the effective electronic temperature ($T_\mathrm{e}$) relaxes toward its equilibrium value through energy exchange with the different degrees of freedom that linearly contribute to $\Pi(\Omega)$. In a more formal description, the  total bosonic function is given by $\Pi(\Omega)$=$\Pi_\mathrm{be}(\Omega)$+$\Pi_\mathrm{SCP}(\Omega)$+$\Pi_\mathrm{lat}(\Omega)$ where $\Pi_\mathrm{be}$ refers to the bosonic excitations of electronic origin at the effective temperature $T_\mathrm{be}$, $\Pi_\mathrm{SCP}$ to the small fraction of strongly-coupled phonons (SCPs) at $T_\mathrm{SCP}$\cite{Perfetti2007} and $\Pi_\mathrm{lat}$  to all other lattice vibrations at $T_\mathrm{lat}$. As the term that couples the rate equations\cite{supporting} for $T_\mathrm{e}$ and $T_\mathrm{b}$ (with $\mathrm{b}$=$\mathrm{be}$,$\mathrm{SCP}$,$\mathrm{lat}$) is $G(\Pi_\mathrm{b},T_\mathrm{b},T_\mathrm{e})/C_\mathrm{b}$ ($G$ being the functional described in \cite{supporting} and $C_\mathrm{b}$ the specific heat of the bosonic population $b$), each subset of the bosonic excitations is characterized by different relaxation dynamics on the femtosecond timescale. The most convenient systems for such an experiment are the hole-doped cuprates close to the optimal dopant concentration needed to attain the maximum critical temperature $T_\mathrm{c}$, in which the total $\Pi(\Omega)$ is maximum\cite{vanheumen2009}. Despite the magnitude of $\Pi(\Omega)$, vertex corrections beyond Eliashberg theory can be reliably neglected\cite{vanheumen2009,LeTacon2011} in this doping regime.
\\

The total bosonic function $\Pi(\Omega)$ is  directly determined by fitting an extended Drude model\cite{supporting} and a sum of Lorentz oscillators accounting for the interband optical transitions in the visible region\cite{Giannetti2011} to the equilibrium dielectric function of optimally doped Bi$_{2}$Sr$_{2}$Ca$_{0.92}$Y$_{0.08}$Cu$_{2}$O$_{8+\delta}$ (Y-Bi2212) high-quality crystals\cite{Eisaki2004} ($T_\mathrm{c}$=96 K), measured at $T$=300 K by conventional spectroscopic ellipsometry\cite{vanderMarel2003}. Assuming a histogram-like form\cite{supporting} (see Fig. S1) and imposing an upper limit of 1 eV, the extracted $\Pi(\Omega)$ is characterized by: i) a low-energy part (up to 40 meV) compatible with the coupling to acoustic phonons\cite{Johnston2011} and Raman-active optical phonons involving $c$-axis motion of the Cu ions\cite{Kovaleva2004}; ii) a narrow, intense peak centered at $\sim$60 meV, attributed to the anisotropic coupling to either out-of-plane buckling and in-plane breathing Cu-O optical modes\cite{Devereaux2004} or bosonic excitations of electronic origin such as spin fluctuations\cite{Dahm2009}; iii) a broad continuum extending up to 350 meV~\cite{Norman2006,Hwang2007,vanheumen2009}, i.e., well above the characteristic phonon cut-off frequency ($\sim$90 meV).
\\

The key point to extend this analysis to non-equilibrium experiments is that the electron self-energy, $\Sigma(\omega,t)$, entering in the calculation of $\sigma(\omega,t)$\cite{supporting}, can be factorized into\cite{Kaufmann1998}:
$$
\Sigma\left(\omega,T\right)=\int^{\infty}_{0}\Pi\left(\Omega\right)L\left(\omega,\Omega,T\right)d\Omega
\label{selfenergy}\;\;\;\;\;\;\;\;\;\;\;\;\;\;(1)
$$
where $L\left(\omega,\Omega,T\right)$ is a material-independent kernel function accounting for the thermal activation of the bosonic excitations and of the QPs. The kernel function $L\left(\omega,\Omega,T_\mathrm{e,b}\right)$= -$2\pi i[N\left(\Omega,T_\mathrm{b} \right)$ +$1/2]$+$\Psi(1/2$+$i(\Omega$-$\Omega')/2\pi T_\mathrm{e})$-$\Psi(1/2$-$i(\Omega$+$\Omega')/2\pi T_\mathrm{e})$, where $\Psi$ is the Digamma function obtained by integrating the Fermi-Dirac functions and $N(\Omega,T_\mathrm{b})$ the Bose distribution at temperature $T_\mathrm{b}$, can be decomposed into different terms depending on the electronic ($T_\mathrm{e}$) and bosonic ($T_\mathrm{b}$) temperatures. The independent variation of $T_\mathrm{e,b}$ is expected to induce different modifications of the dielectric function. Figure 1 shows the expected relative variation of the reflectivity, i.e., $\delta R/R\left(\omega,T\right)$= $[R\left(\omega,T+\delta T\right)$-$R\left(\omega,T\right)]/R\left(\omega,T\right)$, in the quasi-thermal ($\delta T_\mathrm{e}$=$\delta T_\mathrm{b}$$>$0) and non-thermal ($\delta T_\mathrm{e}$$>$0, $\delta T_\mathrm{b}$=0)  scenarios. Phenomenologically, in the first case the reflectivity variation is dominated by the increase of the QP-boson scattering, corresponding to a broadening of the Drude peak, while in the second case the decoupling between the QP and bosonic distributions can be rationalized in terms of a small increase of the plasma frequency without any change in the scattering rate. The difference between the two cases is more significant in the spectral region close to the dressed plasma frequency, $\Omega_\mathrm{p}$$\simeq$1 eV, i.e., an energy scale much higher than the energy scale of the bosonic function.
\\
\begin{figure}
\includegraphics[keepaspectratio, bb= 20 270 550 560, clip,width=0.5\textwidth] {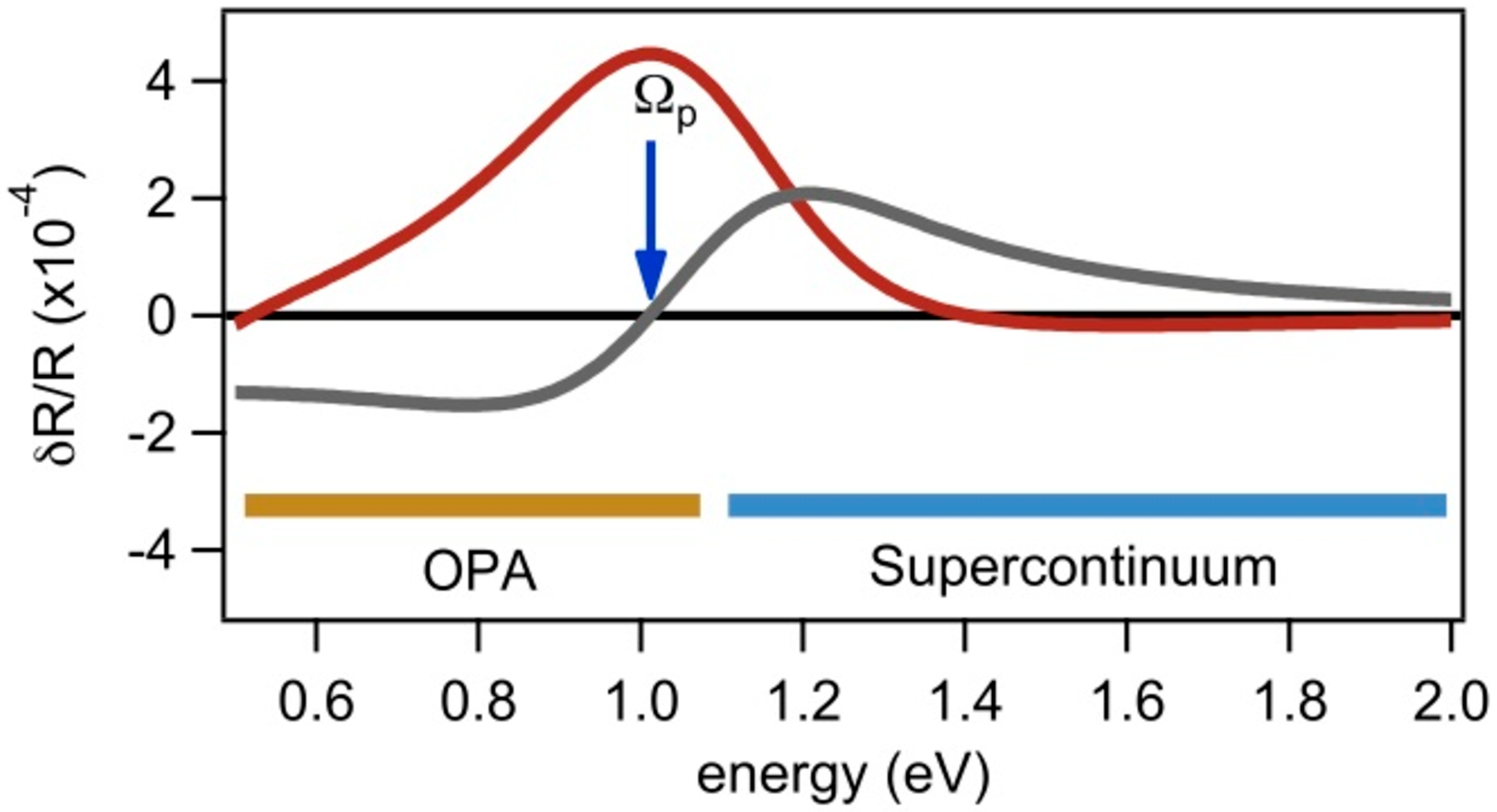}
\caption{Non-equilibrium reflectivity. The relative reflectivity variation $\delta R/R\left(\omega\right)$= $[R\left(\omega,T+\delta T\right)$-$R\left(\omega,T\right)]/R$, calculated for the incremental variations $\delta$T$_{e}$ and $\delta$T$_{b}$ in Eq. 1 and using the $\Pi(\omega)$ obtained from the fit to the equilibrium measurements, is reported in the Fig. S1. The change of sign of $\delta R\left(\omega\right)/R$ in the quasi-thermal scenario (grey line, $\delta$T$_{e}$=$\delta$T$_{b}$=1 K) and the maximum $\delta R\left(\omega\right)/R$ in the non-thermal scenario (red line, $\delta$T$_{e}$=5 K, $\delta$T$_{b}$=0)  coincides with the dressed plasma frequency, $\Omega_{p}$$\sim$1 eV. The spectral regions probed by the Optical Parametric Amplifier (OPA) and supercontinuum techniques\cite{supporting} are marked by the bars with different colors.}
\end{figure}

Time-resolved reflectivity measurements in the 0.5-2 eV photon-energy range \cite{Cilento2010} have been performed at $T$=300 K on the same crystals\cite{Eisaki2004} (OP96) used for the equilibrium optical spectroscopy. The $\delta R/R\left(\omega,t\right)$ two-dimensional matrix is reported in Fig. 2A and B, along with the time-traces at $\sim$1.5 and 0.7 eV photon energies (white curves). After the pump excitation at $t$=0, the temporal dynamics above and below $\Omega_\mathrm{p}$ are very similar, exhibiting a relaxation dynamics of about 200 fs, generally attributed  to the thermalization of electrons with SCPs\cite{Perfetti2007}, and a slower decay on the ps timescale, related to the thermalization with all other lattice vibrations. In Fig. 2C the energy-resolved traces at fixed delay ($t$=100 fs) are reported. Comparing the $\delta R/R\left(\omega\right)$ measured on OP96 to  the relative variation of the reflectivity calculated in the non-thermal and quasi-thermal cases (Fig. 1), we come to the major point of our work: on a timescale faster than the electron-phonon thermalization the electrons are already thermalized with some bosonic excitations participating to $\Pi(\Omega)$. The fast timescale ($\ll$100 fs) of this thermalization implies a very large coupling and a relatively small specific heat. These overall observations strongly suggest that this process involves bosonic excitations of electronic origin. 
\\
\begin{figure}
\includegraphics[keepaspectratio, bb= 30 170 570 680, clip,width=0.65\textwidth] {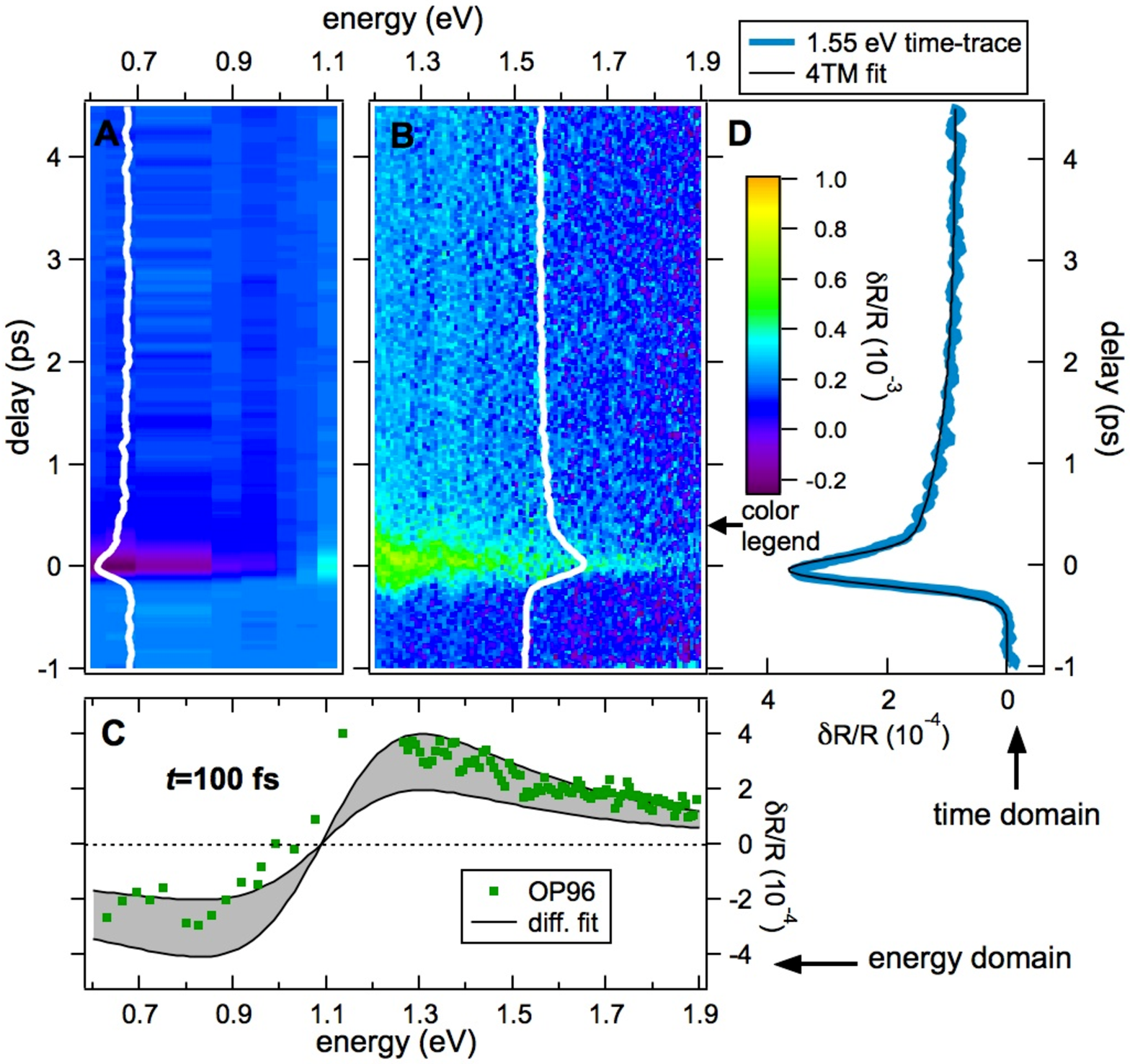}
\caption{Non-equilibrium optical spectroscopy. Time- and frequency- resolved reflectivity data carried out on the OP96 sample at T=300 K in the OPA (A) and supercontinuum (B) probe optical region\cite{supporting}. The white curves represent the time-traces at 1.55 eV and 0.65 eV photon energy. (C) Energy-resolved spectra for OP96, measured at fixed time $t$=100 fs. The black lines are the maximum and minimum $\delta R/R(\omega, t$=100fs), determined through the fitting procedure described in the text and in \cite{supporting} and acounting for the experimental uncertainty in the laser fluence and spot dimensions. Most of the values of $\delta R/R(t$=100fs) measured on OP96 at each wavelength, fall between the two lines. (D) Time-resolved trace for OP96 at fixed 1.55 eV photon energy. The black line is the best fit to the data obtained though the fitting procedure described in \cite{supporting}.}
\end{figure}

The relative strengths of $\Pi_\mathrm{b}(\Omega)$ determine\cite{Allen1987} both the temporal evolution of the temperatures $T_\mathrm{b}$, through the four-temperature model (4TM)\cite{supporting}, and the intensity of the reflectivity variation, through Eq. 1. As a consequence, the simultaneous fit of the calculated $\delta R/R\left(\omega,T_\mathrm{e},T_\mathrm{be},T_\mathrm{SCP},T_\mathrm{lat}\right)$ to the data reported in Fig. 2 in the time- and frequency domain, significantly narrows the phase-space of the parameters of the model, as compared to single-color measurements. The fit procedure\cite{supporting} allows us to unambiguously extract the different contributions to $\Pi(\Omega)$ and to estimate $C_\mathrm{be}$ and $C_\mathrm{SCP}$.  
\\

Figure 3 summarizes the main results of this work. The 10 $\mathrm{\mu}$J/cm$^2$ pump pulse gently increases the electronic temperature by $\delta T_\mathrm{e}$$\sim$2 K. The entire high-energy part and $\sim$46$\%$ of the peak (red areas) instantaneously thermalize with electrons at a temperature $T_\mathrm{be}$$\simeq$$T_\mathrm{e}$. The spectral distribution and the value of the specific heat of these excitations ($C_\mathrm{be}$$<$0.1$C_\mathrm{e}$) demonstrate their electronic origin. On a slower timescale (100-200 fs), the electrons thermalize with the SCPs that represent $\sim$20$\%$ of the phonon density of states ($C_\mathrm{SCP}$=0.2$C_\mathrm{lat}$), but are responsible for $\sim$34$\%$ of the coupling (blue area) in the peak of the bosonic function at 40-75 meV, corresponding to $\sim$17$\%$ of the total bosonic function. Prominent candidates as SCPs are the buckling and breathing Cu-O optical modes. The third and last measured timescale is related to the thermalization with all other lattice modes (80$\%$ of the total) that include all acoustic modes and the IR- and Raman-active modes involving c-axis motion of the Cu ions and provide $\sim$20$\%$ of the coupling (green area) in the peak of $\Pi(\Omega)$.  
\\
\begin{figure}
\includegraphics[keepaspectratio, angle=-90, bb= 90 50 570 800, clip,width=0.8\textwidth] {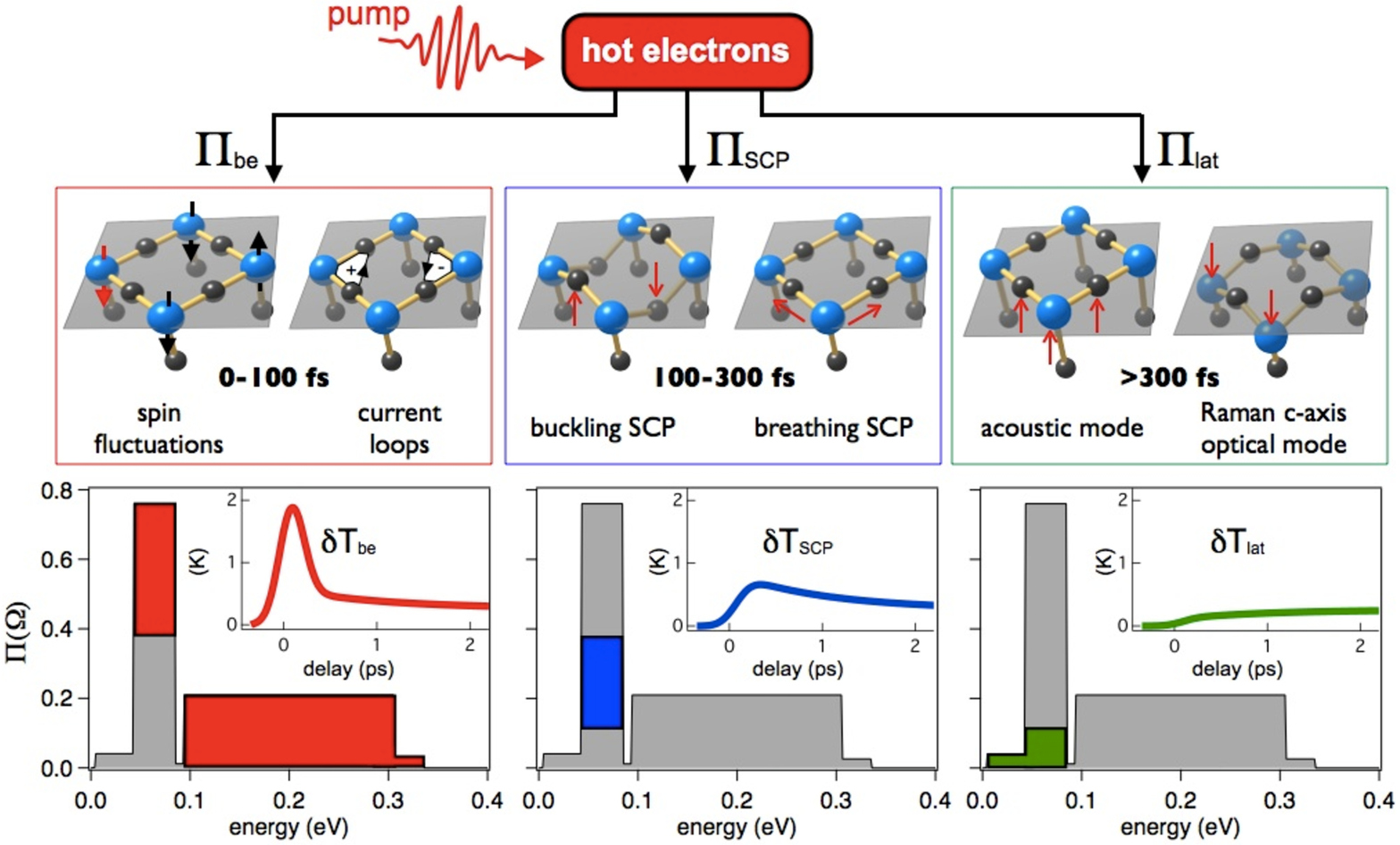}
\caption{Disentangling the contributions to the total bosonic function. The electronic ($\Pi_\mathrm{be}$, red area), strongly-coupled phonons ($\Pi_\mathrm{SCP}$, blue area) and lattice ($\Pi_\mathrm{lat}$, green area) contributions to the total bosonic function are reported. The insets display the temporal evolution of the temperatures $T_\mathrm{be}$ (red line), $T_\mathrm{SCP}$ (blue line) and $T_\mathrm{lat}$ (green line), determined through the 4TM\cite{supporting}. Sketches of the possible microscopic mechanisms at the base of the different contributions to the total $\Pi(\Omega)$ are reported in the top panels.}
\end{figure}

These results have important implications for the identification of the pairing mechanism in cuprates. 
The electron-boson coupling $\lambda_\mathrm{b}$=$2\int\Pi_\mathrm{b}\left(\Omega\right)/\Omega\;d\Omega$ is calculated for each subset b of the bosonic excitations, considering the experimental uncertainties. In the strong-coupling regime ($\lambda_\mathrm{b}$$<$1.5), the critical temperature can be estimated\cite{supporting} through an extended version of the Mc Millan's equation\cite{Allen1975}, containing the logarithmic-averaged frequency ($\tilde{\Omega}$) and the portion of $\Pi_\mathrm{b}(\Omega)$ that contributes to the $d$-wave pairing\cite{Millis1988}. The maximal critical temperature attainable is calculated assuming that each $\Pi_\mathrm{b}(\Omega)$ entirely contributes to the $d$-wave pairing. The coupling with SCPs ($\lambda_\mathrm{SCP}$=0.4$\pm$0.2) is in complete agreement with the values measured on similar materials via different techniques, such as time-resolved photoemission spectroscopy\cite{Perfetti2007}, time-resolved electron diffraction\cite{Carbone2008} and single-color high-resolution time-resolved reflectivity\cite{Gadermaier2010}. Although this value is rather close to the threshold of the strong-coupling regime\cite{Mishchenko2004,Sangiovanni2006}, the small value of $\tilde{\Omega}$ gives $T_\mathrm{c}$=2-30 K, that is far from being able to account for the high-temperature superconductivity of the system. The coupling of electrons with all other lattice vibrations is even smaller in strength ($\lambda_\mathrm{lat}$=0.2$\pm$0.2) and provides an upper bound of the critical temperature of $T_\mathrm{c}$$\simeq$12 K. Finally, the large coupling constant ($\lambda_\mathrm{be}$=1.1$\pm$0.2) and the larger $\tilde{\Omega}$ value of the electronic excitations give $T_\mathrm{c}$=105-135 K and hence accounts alone for the high-critical temperature. We remark that, while $\Pi_\mathrm{SCP}(\Omega)$ and $\Pi_\mathrm{lat}(\Omega)$ are expected to be temperature-independent, $\Pi_\mathrm{be}(\Omega)$ increases as new magnetic excitations emerge when approaching $T_c$, particularly in the pseudogap phase\cite{Li2010}. Therefore, the use of $\Pi_\mathrm{be}(\Omega)$ determined at $T$=300 K to estimate the QP-boson coupling and $T_\mathrm{c}$ underestimates the electronic contribution to the pairing, further supporting our conclusion of a dominant electronic mechanism in the superconductivity of cuprates. All the $\lambda_\mathrm{b}$ values, maximum attainable critical temperatures, and important parameters for each subset of the total bosonic function are reported in the Table S2. 
\\

The measured values of $\lambda_\mathrm{be}$, $\lambda_\mathrm{SCP}$ and $\lambda_\mathrm{lat}$ and the spectral distribution of the bosonic excitations strongly indicate that the antiferromagnetic spin fluctuations\cite{Dahm2009,LeTacon2011} and the loop currents\cite{Varma2006} are the most probable mediators for the formation of Cooper pairs. An isotope effect in the dispersion of nodal QPs has been recently reported on optimally-doped Bi2212, by ARPES measurements\cite{Iwasawa2008}. These data have been carefully analyzed\cite{Schachinger2009}, demonstrating that the nodal isotope effect can be explained assuming that the QP-phonon coupling represents about 10$\%$ of the total contribution of other bosonic excitations. From Figure 3, we estimate that the contribution of SCP is less than 20$\%$ of the total bosonic function. This demonstrates that our results fully explain the observed isotope effect, eventually providing a consistent interpretation of the most important experimental results about the QP-boson coupling in cuprates. 
\\

We remark that our conclusions are rather independent of the assumption of the histogram-like form of $\Pi(\Omega)$ and robust against modifications of the details of the equilibrium dielectric function. In fact, the outcome of this work strongly supports the factorization of the self-energy at $T$=300 K into a temperature-dependent kernel function and the glue function $\Pi(\Omega)$ (see Eq. 1), even under non-equilibrium conditions. Although we do not exclude \textit{a-priori} that the upper limit used in the determination of the bosonic function can hide possible contributions to $\Pi(\Omega)$ even above 1 eV and that the electron-phonon coupling may cooperate in driving the superconducting phase transition, we demonstrate that bosonic excitations of electronic origin are the most important factor in the formation of the superconducting state at high temperatures in the cuprates. 
\\

The present findings pave the way for the investigation of the electron-boson coupling in a variety of complex materials, ranging from transition-metal oxides to iron-based superconductors.
\\

We acknowledge discussions and suggestions from M. Capone and A. Chubukov. F.C., G.C., and F.P. acknowledge the support of the Italian Ministry of University and Research under Grant Nos. FIRBRBAP045JF2 and FIRB-RBAP06AWK3. The crystal growth work was performed in M.G.Õs prior laboratory at Stanford University, Stanford, CA 94305, USA, and supported by DOE under
Contract No. DE-AC03-76SF00515. The work at UBC was supported by the Killam, Sloan Foundation, CRC, and NSERCÕs Steacie Fellowship Programs (A. D.), NSERC, CFI, CIFAR Quantum Materials, and BCSI. D. v.d.M. acknowledges the support of the Swiss National Science Foundation under  Grant No. 200020-130052 and MaNEP.

\bibliography{biblio}

\begin{thebibliography}{33}
\bibitem[{sup()}]{supporting}
\bibinfo{note}{Materials and methods are available as supporting material on
  Science Online.}
\end{thebibliography}

\section*{Materials and methods}
\subsection*{Time-resolved spectroscopy}
The dynamics of the non-equilibrium optical response is probed by ultrashort supercontinuum light pulses obtained by focusing the output beam of a cavity-dumped Ti:sapphire oscillator in a photonic crystal fiber (\textit{25}). The spectral window explored by the white pulses (1.2-2 eV) is extended in the infrared region down to 0.5 eV, using an Optical Parametric Amplifier (OPA), seeded by a 250 kHz amplified  Ti:sapphire oscillator. The pump fluence is always set to 10$\pm$3 $\mu$J/cm$^2$. The temporal resolution of the probe pulses varies with the wavelength from 180 fs at 800 nm to $<$100 fs at 2000 nm. The diameters of the pump and probe beams on the sample are 38$\pm$2 and 20$\pm$3 $\mu$m (Full-Width-at-Half-Maximum).
\subsection*{Critical temperature}
In the strong-coupling formalism, the critical temperature for $d$-wave pairing in a Fermi liquid with pseudopotential $\mu^*$=0 is approximately given by (\textit{27, 28}):
$$
T_{c}=0.83\tilde{\Omega}\exp[-1.04(1+\lambda_b)/g\lambda_b]\;\;\;\;\;\;\;\;\;\;\;\;\;\;(S1) 
$$
where ln$\tilde{\Omega}$=2/$\lambda_b$$\int_0^{\infty}\Pi_b(\Omega)\mathrm{ln}\Omega /\Omega d\Omega$, $\lambda_b$=$2\int\Pi_{b}\left(\Omega\right)/\Omega\;d\Omega$ is the electron-boson coupling constant and $g$$\in$[0,1] is a parameter that accounts for the $d$-wave nature of the superconducting gap. The upper bound $g$=1 corresponds to the case in which $\Pi_b(\Omega)$ entirely contributes to the $d$-wave pairing.

\begin{table}[h]
\caption{The important parameters that have been determined in this work for the pairing mediated by the different kinds of bosonic excitations are reported. The error bars of the values have been evaluated accounting for all the uncertainties in the experimental parameters.}
\begin{center}
\renewcommand{\arraystretch}{1.2}
\centering
\begin{tabular}{lllll}
\\
\multicolumn{1}{c}{Bosonic glue} & \multicolumn{1}{c}{$\lambda$} & \multicolumn{1}{c}{$\tilde{\Omega}$ (meV)} & \multicolumn{1}{c}{$C$} & \multicolumn{1}{c}{T$_{c}$ (K)} \\
\hline \hline
SCPs & 0.4$\pm$0.2 & 60 & 0.2$C_{lat}$ & 2-30 \\
\hline
Lattice & 0.2$\pm$0.2 & 47 & 0.8$C_{lat}$ & 0-12 \\
\hline
\textbf{Electronic} & \textbf{1.1$\pm$0.2} & \textbf{87} & \textbf{$<$0.1$\mathbf{C_e}$} & \textbf{105-135} \\
\hline
Total & 1.7 & 69 & & 137 \\
\end{tabular}
\end{center}
\end{table}
\subsection*{Four temperatures model}
A four-temperatures model can be used to represent the following physical processes: a short laser pulse, with power density (absorbed) $p$, impulsively increases the effective electronic temperature of the electrons with a specific heat $C_{e}$=$\gamma_e$$T_e$ ($\gamma_e$=$\pi^2N_cN(\epsilon_F)k^2_b/3$), $N_c$ being the number of cells in the sample and $N(E_F)$ the density of states of both spins per unit cell). $T_e$ will then relax through the energy exchange with all the coupled degrees of freedom  that linearly contribute to the total $\Pi(\Omega)$=$\Pi_{be}(\Omega)$+$\Pi_{SCP}(\Omega)$+$\Pi_{lat}(\Omega)$.
The rate of the energy exchange among the different populations is given by (\textit{26}):
$$
\frac{\partial T_e}{\partial t}=\frac{G(\Pi_{be},T_{be},T_{e})}{\gamma_eT_e}+\frac{G(\Pi_{SCP},T_{SCP},T_{e})}{\gamma_eT_e}+\frac{G(\Pi_{lat},T_{lat},T_{e})}{\gamma_eT_e}+\frac{p}{\gamma_e T_e}\;\;\;\;\;\;\;\;\;\;\;\;\;\;(S2) 
$$
$$
\frac{\partial T_{be}}{\partial t}=-\frac{G(\Pi_{be},T_{be},T_{e})}{C_{be}}\;\;\;\;\;\;\;\;\;\;\;\;\;\;(S3)\\
$$
$$
\frac{\partial T_{SCP}}{\partial t}=-\frac{G(\Pi_{SCP},T_{SCP},T_{e})}{C_{SCP}}\label{4TM3}\;\;\;\;\;\;\;\;\;\;\;\;\;\;(S4)
$$
$$
\frac{\partial T_{lat}}{\partial t}=-\frac{G(\Pi_{lat},T_{lat},T_{e})}{C_{lat}}\label{4TM4}\;\;\;\;\;\;\;\;\;\;\;\;\;\;(S5) 
$$
where 
$$
G(\Pi_{b},T_{b},T_{e})=\frac{6\gamma_e}{\pi \hbar k^2_b}\int^{\infty}_0d\Omega \Pi_{b}(\Omega) \Omega^2 [N(\Omega,T_{b})-N(\Omega,T_{e})]\;\;\;\;\;\;\;\;\;\;\;\;\;\;(S6) 
$$
with $b$=$be$,$SCP$,$lat$ and $N(\Omega,T)$=($e^{\Omega/k_BT}-1$)$^{-1}$ the Bose-Einstein distribution at the temperature $T$. The specific heat ($C_{SCP}$) of SCPs is proportional to their density of states and is taken as a fraction $f$ of the total specific heat, i.e., $C_{SCP}$=$fC_{lat}$.
We pinpoint that the 4TM used does not contain any term for the anharmonic decay of SCP (\textit{16}), that directly couples Eqs. S4 and S5. Including this term, which is of difficult evaluation, would eventually lead to a smaller value of $\Pi_{lat}(\Omega)$ and a larger value of $\Pi_{be}(\Omega)$.\\

The four temperature model relies on the assumption that quasi-thermal electronic and bosonic distributions, thermodynamically defined by the effective temperatures $T_e$ and $T_b$ (with $b$=$be$,$SCP$,$lat$), are rapidly established. This corresponds to assume that the electron-electron scattering time ($\tau_{e-e}$) is much smaller than the electron-phonon scattering time ($\tau_{e-ph}$). The evaluation of $\tau_{e-e}$ in strongly-correlated systems and in cuprates, in particular, is a subtle and open problem, since the applicability of arguments based on the Fermi-liquid theory is questionable (\textit{34}). In Refs. (\textit{31, 35}) an analytical approach, based on the Boltzmann equation and free from any quasi-equilibrium approximation, has been used to extract the electron-phonon coupling from single-color time-resolved measurements, obtaining results fully compatible with our conclusions ($\lambda$$\sim$0.5 in La$_{1.85}$Sr$_{0.15}$CuO$_{4}$ and $\sim$0.25 in YBa$_{2}$Cu$_{3}$O$_{6.5}$). Using the solution of the Boltzmann equation, the authors of (\textit{31}) demonstrated that the quasi-thermal models overestimate $\lambda_{e-ph}$ of a factor 8/5, in the extreme case $\tau_{e-e}$/$\tau_{e-ph}$$\rightarrow$$\infty$. In our case, the fit of the 4 temperature model to the time-resolved data gives $\lambda_{SCP}$=0.37 that, divided by 8/5, becomes 0.23, well within the error bars ($\lambda_{SCP}$=0.4$\pm$0.2) reported in our work. In conclusion, the large error bars associated to the values of $\lambda_{SCP}$ and $\lambda_{lat}$ include: i) the experimental uncertainty in the pump fluence; ii) the possibility of adding to the 4TM a term that accounts for the anharmonic coupling of $SCP$ and the lattice; iii) the possibility of the failure of the 4TM in the extreme case $\tau_{e-e}$$\gg$$\tau_{e-ph}$.\\
\subsection*{Equilibrium dielectric function}

Figure S1 reports the \textit{ab}-plane reflectivity, as measured at 300 K on optimally doped Bi$_{2}$Sr$_{2}$Ca$_{0.92}$Y$_{0.08}$Cu$_{2}$O$_{8+\delta}$ (Y-Bi2212) high-quality crystals (T$_{c}$=96 K), by conventional spectroscopic ellipsometry (\textit{21}). The dielectric function has been obtained by applying the Kramers-Kronig relations to the reflectivity for 50 cm$^{-1}$$<$$\omega$/2$\pi c$$<$6000 cm$^{-1}$ and directly from ellipsometry for 1500 cm$^{-1}$$<$$\omega$/2$\pi c$$<$36000 cm$^{-1}$. This combination allows a very accurate determination of $\epsilon(\omega)$ in the entire combined frequency range. Due to the off-normal angle of incidence used with ellipsometry, the  \textit{ab}-plane pseudo-dielectric function had to be corrected for the  \textit{c}-axis admixture.

The solid black line in the Fig. S1 is the fit to the data of an extended Drude model, describing the strong coupling of the free carriers with the bosonic excitations, and a sum of Lorentz oscillators accounting for the interband optical transitions in the visible region (\textit{19}). 
\begin{figure}
\includegraphics[keepaspectratio, bb= 30 220 550 630, clip,width=0.5\textwidth] {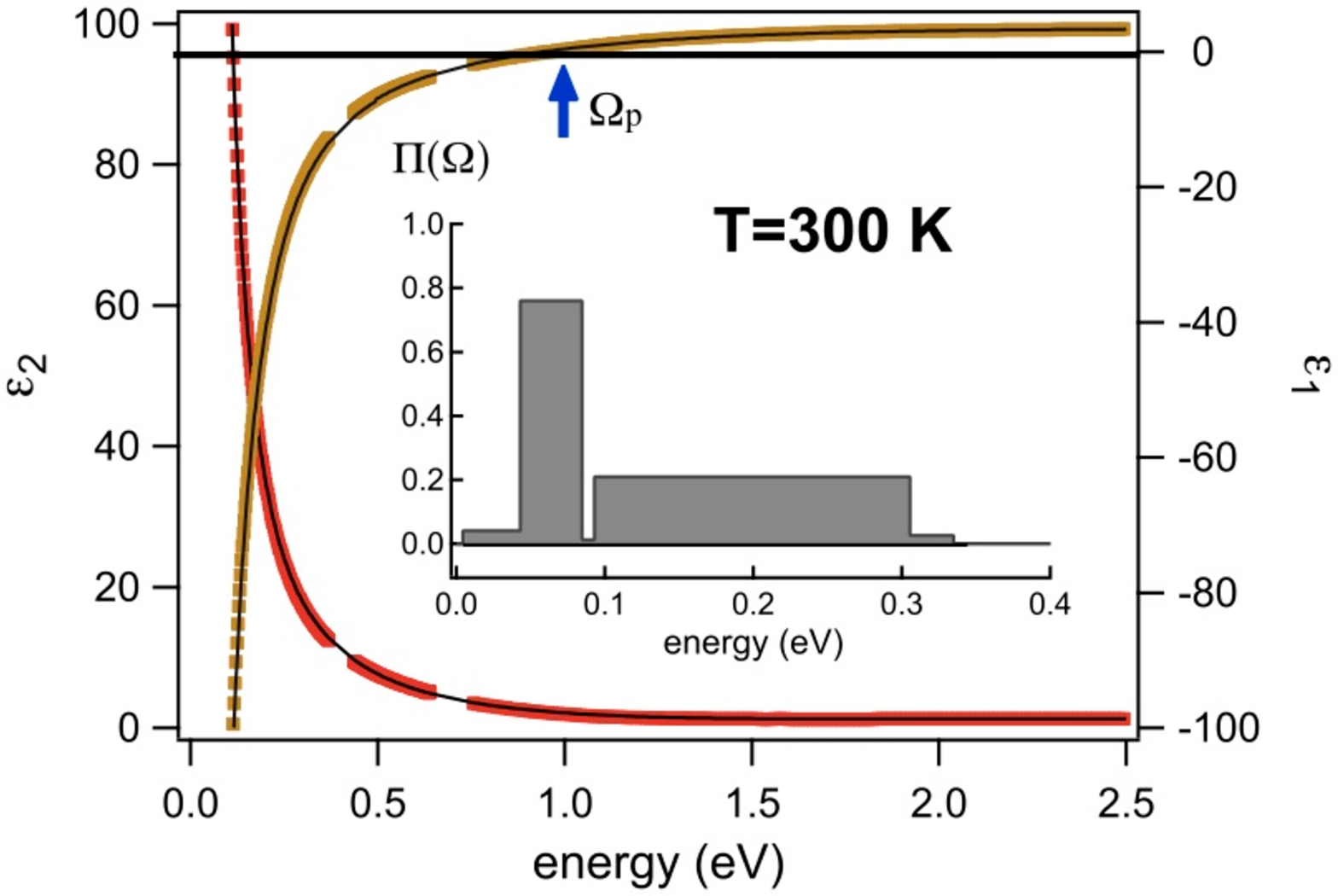}
\caption{The real part ($\epsilon_1$, right axis) and the imaginary part ($\epsilon_2$, left axis) of the equilibrium dielectric function of an optimally doped Y-Bi2212 crystal, measured over a wide energy range at T=300 K, are reported. The inset shows the bosonic function $\Pi(\Omega)$, extracted from the optical spectra. The blue arrow indicates the dressed plasma frequency $\Omega_{p}\sim1$ eV. The black curves are the best fit of the extended Drude model to the data.}
\end{figure}

Within the extended Drude model, a simplified expression of $\sigma(\omega)$ is obtained (\textit{32}) in terms of the one-particle self-energy, by omitting the vertex corrections (Migdal approximation):
$$
\sigma\left(\omega,T\right)=\frac{\omega^{2}_{p}}{i4\pi\omega}\int_0^{\infty}\frac{f\left(\omega+\epsilon,T\right)-f\left(\epsilon,T\right)}{\omega-\Sigma\left(\omega+\epsilon,T\right)+\Sigma^{\ast}\left(\epsilon,T\right)}\;d\epsilon\;\;\;\;\;\;\;\;\;\;\;\;\;\;(S7) 
$$
where $\omega_{p}$ is the plasma frequency and $f\left(\epsilon,T\right)$ the Fermi distribution at temperature $T$. $\Sigma\left(\omega\right)$ and $\Sigma^{\ast}\left(\omega\right)$ are the electron and hole self-energies (obtained by averaging $\Sigma\left(\omega,\textbf{k}\right)$ over the Fermi surface and assuming a constant density of states)~(\textit{24}):
$$
\Sigma\left(\omega,T\right)=\int^{\infty}_{0}\Pi\left(\Omega\right)L\left(\omega,\Omega,T\right)d\Omega \;\;\;\;\;\;\;\;\;\;\;\;\;\;(S8) 
$$
where $L\left(\omega,\Omega,T\right)$ is a material-independent kernel function accounting for the thermal excitations of the glue and the QPs. 

The normal-incidence reflectivity $R(T,\omega)$ is calculated from Eq. S7 through the relation: 
$$
R(T,\omega)=\left| \frac{1-\sqrt{\epsilon(T,\omega)}}{1+\sqrt{\epsilon(T,\omega)}}\right|^2\;\;\;\;\;\;\;\;\;\;\;\;\;\;(S9) 
$$
$\Pi(\Omega)$ is  directly determined fitting the reflectivity obtained from Eq. S7-S9 to the optical spectra reported in the Fig. S1. 
\subsection*{Differential fitting procedure}
The relative reflectivity variation, i.e. $\delta R/R(\omega,t)$, is a functional of $\Pi_b(\Omega)$ ($b$=$be$,$SCP$,$lat$) (see Eqs. S7, S9) and of $C_b$, that are the parameters determining the temporal evolution of $T_b$ through the four-temperature model described in the previous section of the Methods. The different contributions to $\Pi(\Omega)$ are extracted fitting $\delta R/R(\omega,t)$ to the time- and frequency-resolved reflectivity data, reported in Figure 2. 

Considering that the energy distribution of phonons is limited to $<$90 meV, we assume that, for $\Omega$$>$90 meV, $\Pi(\Omega)$$\simeq$$\Pi_{be}(\Omega)$. Within this assumption, the functional dependence of $\delta R/R(\omega,t)$ on $\Pi_b(\Omega)$ is simplified as a parametric dependence on the coefficients $p_b$, where $\Pi_{be}(\Omega)$=$p_{be}$$\Pi(\Omega$$<$90 meV)+$\Pi(\Omega$$>$90 meV), $\Pi_{SCP}(\Omega)$=$p_{SCP}$$\Pi(\Omega$$<$90 meV) and $\Pi_{lat}(\Omega)$=$p_{lat}$$\Pi(\Omega$$<$90 meV).

Considering the constraints given by the relations $\delta R/R(\omega,t)$=$F(p_{be},p_{SCP},p_{lat},C_{be},$ $C_{SCP},C_{lat};\omega,t)$ ($F$ being the generic function expressing the dependence on the parameters contained in Eqs. S7-S9) and $\Pi(\Omega)$=$\sum_b\Pi_b(\Omega)$ ($\Pi(\Omega)$ being the total glue function reported in the Figure 2), the free parameters of the fitting are four. From the fit to the rise time of the time-resolved signal, it is possible to obtain an upper limit to the value of the specific heat of the electronic excitations, i.e., $C_{be}$$\leq$0.1$C_e$. Fixing the values $C_e$/$T_e$=$\gamma_e$=10$^{-4}$ Jcm$^{-3}$K$^{-2}$ and $C_{lat}$=2.27 Jcm$^{-3}$K$^{-1}$, we are able to simultaneously perform the fit to the data reported in the Figure S1 in the time and frequency domain with only two free parameters, i.e. $C_{SCP}$ and any of the $p_b$. Furthermore, the small influence of the variations of $C_{SCP}$ on the fit results (i.e. $\partial F/\partial C_{SCP}$$\ll$$\partial F/\partial p_{b}$), significantly narrows the parameters phase-space of the model, allowing to unambiguously haul out the different contributions to $\Pi(\Omega)$. 

To carry out a quantitative and reliable analysis of the results, both the finite lateral dimension of the probe beam and the inhomogeneous excitation of the pump pulse, related to the finite penetration depth, have been accounted for. Considering a pump pulse of energy $P$ and a probe pulse with gaussian shapes in the radial coordinate, the measured reflectivity variation is given by:
$$
\frac{\delta R}{R}\propto \int_0^{\infty}\frac{1}{\sigma^2_{pr}}e^{-r^2/\sigma^2_{pr}}\frac{P}{\sigma^2_{pu}}e^{-r^2/\sigma^2_{pu}}rdr=\frac{P}{2(\sigma^2_{pr}+\sigma^2_{pu})}\;\;\;\;\;\;\;\;\;\;\;\;\;\;(S10) 
$$
$\sigma_{pr(pu)}$=FWHM/2$\sqrt{\mathrm{ln2}}$ being the lateral size of the probe(pump) beam. As compared to the reflectivity variation calculated for a homogeneous excitation with a surface density of 4$P$/FWHM$^2$, the measured $\delta R/R(\Omega)$ is suppressed of a factor ln2$\cdot$FWHM$_{pu}$$^2$/(FWHM$_{pu}$$^2$+FWHM$_{pr}$$^2$).

The role of the finite penetration depth of the pump pulse ($d_{pu}$=160 nm @ 1.55 eV) is accounted for  numerically calculating $\delta R/R(\Omega)$ through a transfer matrix method, when a graded index of the variation of the refractive index $n$ with exponential profile, i.e. $\delta n$=$\delta n_0$$e^{-z/d_{pu}}$ ($d_{pu}$ being the penetration depth of the pump pulse), is assumed.
\subsection*{Supplementary References}
34. D. Fournier et al. \textit{Nature Physics} \textbf{6}, 905 (2010).

35. V.V Kabanov and A.S. Alexandrov, \textit{Phys. Rev. B} \textbf{78}, 174514 (2008).

\end{document}